# Inter-diffusion, melting and reaction interplay in Ni/4H-SiC under excimer laser annealing

**Salvatore Sanzaro[1], Corrado Bongiorno[1], Paolo Badalà[2], Anna Bassi[2], Giovanni Franco[2], Patrizia Vasquez[2], Alessandra Alberti[1,*] and Antonino La Magna[1]**

[1]CNR-IMM, Zona Industriale Strada VIII 5, 95121 Catania, Italy
[2]STMicroelectronics, Zona Industriale Stradale Primosole 50, 95121 Catania, Italy

*(\*) Corresponding author:* alessandra.alberti@imm.cnr.it
DOI: https://doi.org/10.1016/j.apsusc.2020.148218

*Abstract*

We investigated the complex interaction between a nickel layer and a 4H-SiC substrate under UV-laser irradiation since the early stages of the atomic inter-diffusion. An exhaustive description is still lacking in the literature. A multimethod approach based on Transmission Electron Microscopy, Energy Dispersive Spectroscopy and Diffraction (electron and X-ray) techniques has been implemented for a cross-correlated description of the final state of the contact after laser irradiation. They detailed the stoichiometry and the lattice structure of each phase formed as well as the Ni/Si alloy profile along the contact for laser fluences in the range 2.4-3.8 J/cm$^2$. To make a bridge between process conditions and post-process characterizations, time dependent ultra-fast phenomena (laser pulse ≈160ns), such as intermixing driven melting and Ni-silicides reactions, have been simulated by a modified phase fields approach in the proper many-compounds formulation. We disclose that raising laser fluence has multiple effects: 1) the thickness of the silicide layer formed at the interface with 4H-SiC expands; 2) the silicon content into the reacted layer increases; 3) a $Ni_2Si$ phase is promoted at the highest fluence 3.8 J/cm$^2$; 4) silicon atomic diffusion into a topmost residual nickel layer occurs, with the Ni/Si ratio increasing towards the contact surface.

1. **Introduction**

Nowadays silicon carbide, especially its 4H-SiC polymorphism [1], represents one of the key materials to fabricate low ON-resistance ($R_{ON}$) and high-power devices [2,3,4]. This technology requires a wafer thinning in the range 110-350 µm in order to reduce the $R_{ON}$ [5]. As a consequence of the wafer thinning, newly conceived thermal treatments are needed for back-side metallization processes that are able to preserve the devices at the front-side of the wafer [3,6]. Knowledge gained in Si-based technologies has been moved to corresponding SiC-based devices and indeed nickel is widely applied in metallization schemes [7,8]. A valid method to induce the formation of Ni-based ohmic contact on 4H-SiC is laser annealing (LA), a viable alternative to Rapid Thermal Annealing (RTA) that has been already extensively used for consolidated Si-based technologies [9,10,11,12,13,14]. In particular, the pulsed UV-lasers typology, offering a suitable penetration depth and the consequent localized annealing at the Ni-side, are applicable for this scope. Despite this approach has been already applied as reported in literature [3,6], a clear understanding of the mechanisms governing the silicide formation on 4H-SiC under laser irradiation is still lacking since the interaction between Ni, Si and C is very complex. Indeed, several interrelated phenomana come into play such as inter-diffusion of species [15,16,17,18,19,20,21], silicide reactions [22,23,24,25,26,27,28] and eventually melting processes [29].

Nickel is a strongly active species and even into SiC matrixes intermixing can be activated at low temperature (*e.g.* > 500°C). Nickel atoms indeed participate to weaken the strong chemical bond between Si and C at the Ni/4H-SiC interface region thus promoting the inter-diffusion of Ni, Si and C; moreover, relevant selective silicon release from SiC is expected also for a pulsed annealing process at high temperatures(~1500 K) in the timescale of hundred ns [15,16,17,18,19,20,21]. While diffusing, Ni and Si could react with the formation of different silicide phases whose stoichiometry depends on the relative amount of the species and on the local temperature. They arise from the thermodynamic equilibrium and the kinetic conditions during the annealing process and during the cooling ramp down [22,25,26,27,28]. Rum-up and down are likewise important in time-dependent temperature fields that are also locally changeable. On this basis, laser pulse and fluence are crucial parameters whose role must be still fully understood [29].

In a previous work [29], we provided an initial description of a Ni-silicided ohmic contact after UV-laser irradiation as a function of the initial Ni layer thickness. Appling also an approximated modeling framework, we emphasized the main role of Ni melting and heat diffusion in relationship with the in-depth thermal field. Here we want to disclose the role of atomic inter-diffusion, reaction and melting phenomena as a function of the laser fluence in order to entail a general paradigm to predict the contact composition and morphology for the optimization of the metallization scheme.

In this framework, the formation of Ni-based contacts on 4H-SiC is studied by applying 1-shot excimer laser irradiation with pulse duration 160 ns at a wavelength of λ = 308 nm at different fluence, 2.4, 3.2 and 3.8 J/cm$^2$, on 100 nm-thick nickel layers. Each treated sample is compared with the sample in as deposited condition used as a reference. The morphological, structural and chemical characterizations are done using different techniques: Transmission Electron Microscopy (TEM) for morphological and local structural analysis, X-Ray Diffraction (XRD) for average structural and compositional analysis, Scanning-Transmission Electron Microscopy (STEM) combined with Energy Dispersive X-Ray spectroscopy for structural and chemical investigation.

A comprehensive modelling is here applied to fully describe the material kinetic during the pulsed heating. Indeed, laser process simulations have been performed by embodying the full Ni-Si reaction path and the dependency of the thermal and optical parameters on the local atomic composition during the inter-diffusion of the atomic species in order to predict the dynamic evolution of the materials under irradiation.

With our systematic and multimethod investigation, we shed light into the complex mechanisms governing the Ni/4H-SiC interaction under UV irradiation since the early stages with predictive valence for future technological approaches to optimize the contact composition and structure. This detailed description is still lacking in the literature.

## 2. Experimental setup

The backside contact has been made on highly doped n-type 4H-SiC substrates, thinned at 110 µm by mechanical grinding. Ni layers, 100 nm-thick, have been deposited on grinded C-side 4H-SiC surfaces by DC sputtering in Ar ambient at a base pressure of $1 \times 10^{-3}$ mbar. Since the silicide reaction is sensitive to impurities at the interface between Ni and substrate [12,30,31], an *in-situ* sputter etching step has preceded Ni deposition in order to ensure a proper interface cleaning. The Ni surfaces have been then irradiated by means of an excimer laser with a wavelength of 308 nm and with several energy densities (fluences Φ) in the range 2.4 - 3.8 J/cm$^2$. The spot size in 10mmx10mm. Structure and composition of the silicide contact have been studied by XRD analysis using a Bruker AXS D8 DISCOVER diffractometer working with a Cu-K$_\alpha$ source and a thin film attachment, by TEM and STEM coupled with EDX using a JEOL-JEM microscope working at 200 keV. A customized multi-elements (Ni monomer, Si monomer, C monomer, C cluster) and multi-compounds (4H-SiC and silicide $Ni_xSi_y=\{NiSi_2, NiSi, Ni_3Si_2, Ni_2Si, Ni_{31}Si_{12}, Ni_3Si\}$) phase fields model [32] has been integrated and calibrated in the LIAB computation tool (see Ref. [33] for the detailed presentation of the approach). This model, which considers the impact of the local composition on the thermal and optical parameters (including melting temperature and latent heat) has been used for the laser annealing simulation of the experimental systems and processes (time dependent pulse shape and fluences).

## 3. Results and discussions

### 3.1 100 nm-Ni/4H-SiC annealed @2.4 J/cm$^2$

#### 3.1.1 Structural and local compositional analysis by TEM and EDX

We studied the morphology and the local structure of the Ni layer after irradiation with 1 shot laser at 2.4 J/cm$^2$ for 160 ns using TEM analyses. After LA, we observed an increment of the thickness of the layer by 22 nm. Moreover, we found that it is made of big grains with defects such as inclusions and twins as shown in **Fig. 1a**. In **Fig. 1b**, the TEM image in dark field shows the presence of some isolated C-clusters at 5-7 nm away from the 4H-SiC substrate (a few small C-clusters can be rarely found up to ~20 nm). The magnification of the layer in **Fig. 1c** (see red square in Fig. 1a) shows with more detail the difference of the morphology between the layer at the interface with 4H-SiC, ~10-20 nm thick, and the rest of the Ni layer. In this latter, we observed twinned defects extending up to the interfacial layer and typical diffraction patterns with spots arranged in a cubic structure as in the Ni lattice, as shown in the Selected Area Electron Diffraction (SAED) reported in **Fig. 1d**.

We performed STEM assisted by EDX analyses in order to obtain the local chemical spatial distribution for each species of interest, *i.e.* Ni, Si and C. Then, **Fig. 1e** shows the EDX spectrum related to the selected light blue area (~7 nm thick at the interface) of the STEM image reported as inset. Here, from the quantitative analysis we found that the Ni:Si ratio is ~3:1 and then the silicide phase formed in the interfacial layer with 4H-SiC is ascribable to Ni$_3$Si that is a cubic phase. Other details of the chemical composition in the rest of the Ni layer are shown in **Fig. 1f**. By STEM analysis we indeed probed a region 100 nm thick as indicated in the selected yellow area in Fig. 1a and in inset Fig. 1e (in this way the surface is exclude because the glue used for the sample preparation could distort the C concentration), in order to obtain the atomic distribution profile of the species of interest (Ni, Si, C). Thus, we found a first layer (~10-20 nm) with composition in agreement with what observed in the Fig. 1c and moreover, we found that the atomic concentration of Si gradually decreases while approaching the sample surface. **Fig. 1g** shows the trend of the Ni:Si ratio along the layer thickness. Here, we notice a growing trend of the Ni:Si ratio, with values that are ~3:1 in proximity of the interface whilst, going toward the surface, the Ni:Si ratio progressively increases describing a Ni matrix with Si diffused inside.

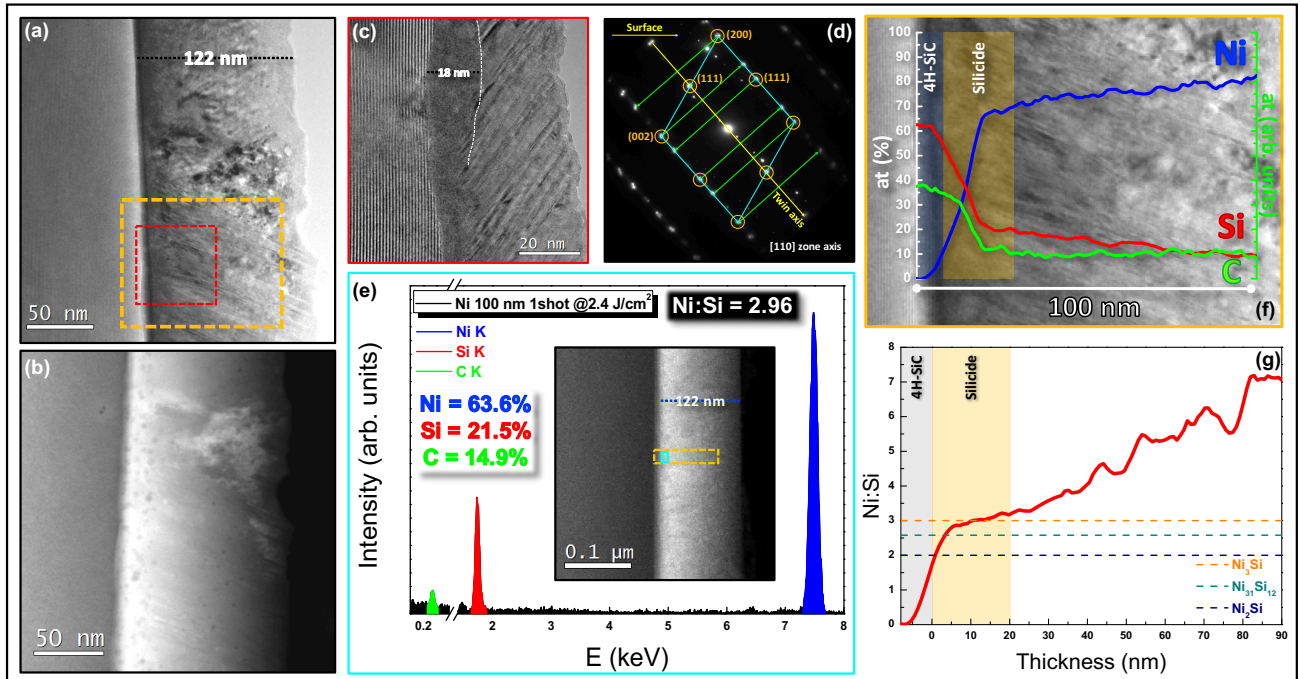

**Fig. 1:** cross-sectional TEM images **(a)** in bright field and **(b)** in dark field of the Ni sample (122 nm) laser annealed 1 shot at 2.4 J/cm$^2$ for 160 ns. **(c)** Magnification of the red panel in the Fig. (a) relative a thin part of the Ni layer in order to show the difference of the Ni layer at the interface with the substrate. **(d)** SAED image taken on the topmost cubic layer having [111] twins (rotation by 180° around the twin axis). **(e)** EDX spectrum of the light blue panel in the STEM image in inset. **(f)** EDX atomic distribution profile in the region inside the yellow panel in the Fig. (a) and in inset of Fig. (e) through the layer. **(g)** Ni:Si ratio profile from the interface with 4H-SiC through the entire layer. An interfacial layer, 15-20 nm-thick, is formed with silicide composition expected as indicated by the dashed lines in the Fig. Above this interface, Si atoms diffuse into the Ni layer as shown by the red curve.

### 3.1.2 Identification of phases by XRD

XRD analysis is performed in different scan configurations, symmetric and grazing incidence, in order to get structural information before (black line) and after (red line) laser annealing. From the acquisition in symmetric configuration and the related rocking curve (see **Fig. 2a** and **c**, respectively), we deduced that the reference sample is highly textured and the texturing is along the [111] direction of the face-centered-cubic Ni structure. After 1-shot laser annealing at 2.4 J/cm$^2$, the material maintains the same preferential orientation along the [111] direction even if the main (111) peak is larger than in the reference sample, as shown in **Fig. 2b**. In particular, this broadening is asymmetric toward lower 2ϑ values, and this is accounted by the presence of a second peak at 2ϑ = 44.266°. All the details on those and following deconvoluted peaks are found in **Table 1**. The extra-peak in Fig. 2b is attributed to Si diffusion into the Ni matrix that increases the d-spacing (d = interplanar distance; Δd/d in Table 1 is referred to pure nickel). Moreover, the crystallite size (the size of domains with high crystallographic order) associated to this Ni-Si alloy is smaller than the one extracted by the Ni peak (21 nm) denoting a structural change of the starting metallization layer. Table 1 also reports the ratio between the areas under the Ni and Ni+Si peaks, providing a qualitative comparison between the two materials, even though it does not consider difference arising from the scattering factors (likely lower in the mixed phase). To have information on the spread of the main growth axis, we performed rocking curves along the main (111) peak as shown in **Fig. 2c**. Here, we estimated a higher Full Width at Half Maximum (FWHM) for the laser annealed sample with respect to the reference (the peak is also less intense) and this means that the sample has partially lost its starting structural order.

From the XRD analysis performed in grazing incidence configuration (see **Fig. 2d**), we concluded that the annealed Ni sample is noticeably different with respect to the reference sample. In fact, after the laser annealing, the sample shows two main peaks, (111) and (200), that are not present in the reference. This is because the reference Ni layer is highly textured with preferential (111)-growth planes. As shown in the corresponding magnification image in **Fig. 2e**, those contributions in the laser treated sample are symptomatic of the formation of a cubic β-Ni$_3$Si phase, in full agreement with the previous compositional EDX local data. The broadening of the (200) peak at 2ϑ = 52.205° likewise accounts for this phase formation with nanosized-grains (~6 nm in diameter). Further details are provided by the deconvoluted contributions, especially the ones located under the first main peak, that describe the presence of three components: Ni+Si, Ni and *β*-Ni$_3$Si (see Table 1 for all the parameters). As a main information, the lattice spacing associated to the (111) peak of the Ni+Si alloy phase is slightly expanded with respect to the one of the pure Ni layer by an amount Δd/d = 0.6%.

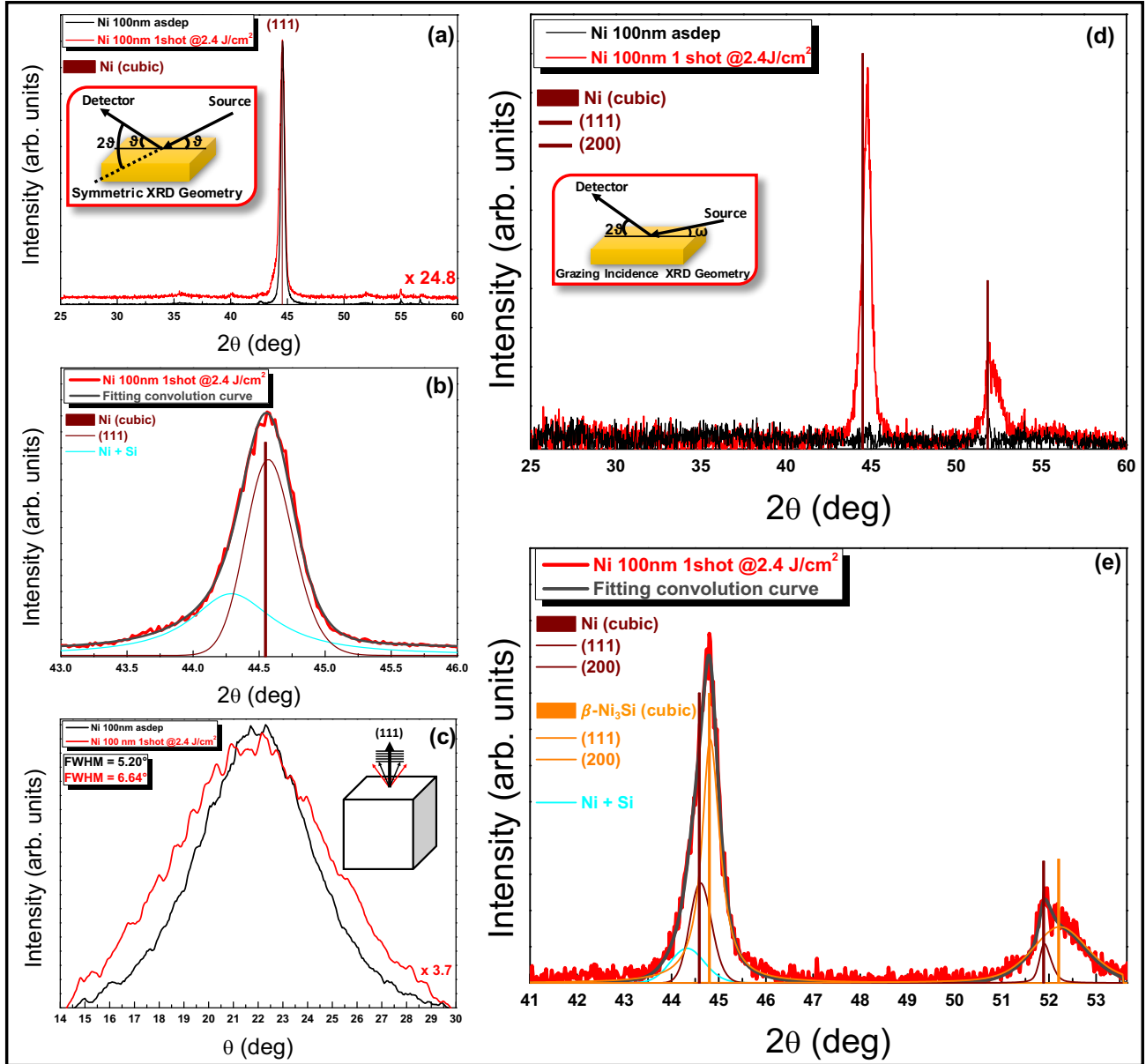

**Fig. 2**: **(a)** normalized XRD patterns in symmetric configuration of the Ni sample laser annealed at 2.4 J/cm$^2$ for 160 ns (**black** line) and in as deposited condition (**red** line). **(b)** Magnification and deconvoluted contributions in the range around the main (111) peak and **(c)** normalized Rocking curves of the treated vs the untreated samples. **(d)** XRD patterns in grazing incidence configuration of the layers after laser annealing at 2.4 J/cm$^2$ for 160 ns (**black** line) and in as deposited condition (**red** line). **(e)** Magnification and deconvoluted contributions of the treated sample in the range around the main (111) and (200) peaks acquired in grazing incidence configuration. The red patterns in (a) and (c) are multiplied by a constant. Together with pure Ni contributions, deconvoluted components of $\beta$-Ni$_3$Si and a Ni+Si mixed alloy are found in the patterns.

**Table 1:** XRD parameters in symmetric and grazing incidence configurations related to the data in Fig. 2 (@2.4 J/cm$^2$) and reporting: position (2ϑ) and interplane lattice distances (d) for each peak; Δd/d is the d variation normalised to the reference Ni peak; Full Width at Half Maximum (FWHM) of each peak; crystallite size (cry. size) for each peak component; area under each deconvoluted peak normalized to the overall area within the angular range reported in column 3 (realtive area %). An instrumental discrepancy between symmetric and grazing peak position of 0.04° is accounted in the reported values. A Ni+Si contribution is emerging from the deconvolution procedure in both geometries with an associated interplane parameter d slightly larger than in the reference untreated nickel layer. A nanostructured $\beta$-Ni$_3$Si phase is formed after laser annealing.

| XRD deconvolution - Ni 100nm @2.4 J/cm² | | | | | | | | | | |
|---|---|---|---|---|---|---|---|---|---|---|
| XRD Geometry | Phase | Peak range (in 2ϑ) | Plane | Position | | Δd/d (%) | FWHM (°) | Cry. Size (nm) | Area (arb. units) | | Relative area (%) |
| | | | | 2ϑ (°) | d (Å) | | | | Relative | Total | |
| Symmetric | Ni + Si | 43° - 46° | - | 44.266 | 2.045 | 0.60 | 0.689 | 12 | 1.5 | | 44.5 |
| | Ni (cubic) | | (111) | 44.548 | 2.033 | - | 0.385 | 21 | 1.9 | | 55.5 |
| Grazing Incidence | Ni + Si | 43° - 46° | - | 44.306 | 2.044 | 0.60 | 0.791 | 10 | 5.7 | 48 | 11.8 |
| | Ni (cubic) | | (111) | 44.588 | 2.031 | - | 0.503 | 16 | 11 | | 22.4 |
| | $\beta$-Ni$_3$Si (cubic) | | (111) | 44.802 | 2.022 | - | 0.397 | 20 | 32 | | 65.8 |
| | Ni (cubic) | 50.5° - 53.5° | (200) | 51.887 | 1.762 | - | 0.225 | 37 | 3.4 | 28 | 12.3 |
| | $\beta$-Ni$_3$Si (cubic) | | (200) | 52.205 | 1.752 | - | 1.317 | 6 | 24 | | 87.7 |

### 3.2 100 nm-Ni/4H-SiC annealed @3.2 J/cm²

#### 3.2.1 Structural and compositional analysis by TEM and EDX

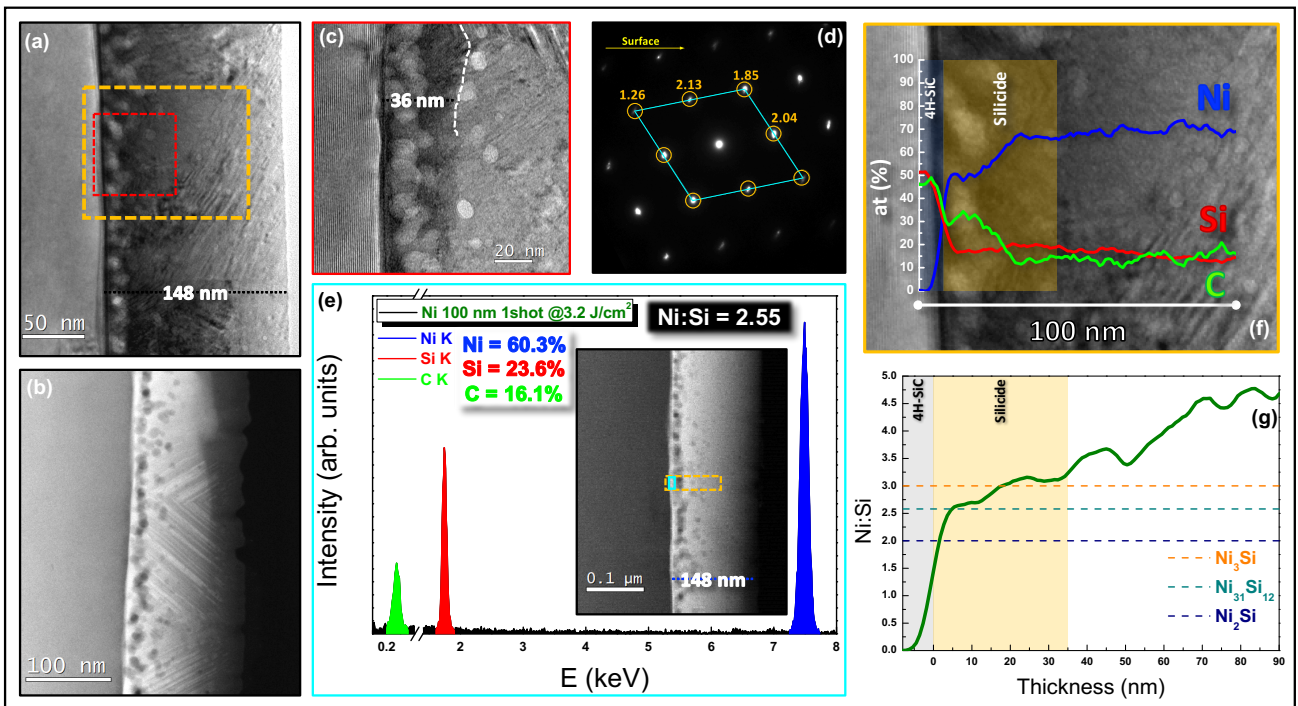

**Fig. 3:** cross-sectional TEM images **(a)** in bright field and **(b)** in dark field of the Ni sample (148 nm) laser annealed 1 shot at 3.2 J/cm² for 160 ns. **(c)** Magnification of the red panel in the Fig. (a) relative a thin part of the Ni layer in order to show the difference of the Ni layer at the interface with the substrate. **(d)** SAED image taken at the interfacial layer that testifies silicide reaction in a not-cubic phase. **(e)** EDX spectrum of the light blue panel in the STEM image in inset. **(f)** EDX atomic distribution profile in the region inside the yellow panel in the Fig. (a) and in inset of Fig. (e) through the layer. An accumulation of C atoms close to the silicide/4H-SiC interface is depicted by the green curve. **(g)** Ni:Si ratio profile from the interface with 4H-SiC through the entire layer. An interfacial layer, 30-35 nm-thick, is formed with silicide composition expected mostly between $\beta$-Ni$_3$Si and $\gamma$-Ni$_{31}$Si$_{12}$ as indicated by the dashed lines in the Fig. Above this interface, Si atoms diffuse into the Ni layer as shown by the green curve.

The same characterizations done for the sample annealed at 2.4 J/cm$^2$ are carried out for the sample 1-shot laser-annealed at 3.2 J/cm$^2$ for 160 ns. Then, in this case, after LA we observed a further increment of the interfacial layer thickness compared with the previous case (22 nm at 2.4 J/cm$^2$ vs. 48 nm at 3.2 J/cm$^2$). Nonetheless, the uppermost part of the layer maintains the same morphological properties, *i.e.* big grains with defects such as inclusions and twins as shown in **Fig. 3a**. From the TEM image in dark field in **Fig. 3b**, we found that the C-clusters increased in number and size; they are still confined within the first 5-7 nm from the substrate even if some C-clusters are found up to ~50 nm from the interface. The magnification shown in **Fig. 3c** is relative to the red panel selected in Fig. 3a. Here, we can observe with more detail the difference of the morphology between the layer at the interface with the substrate, ~30-40 nm thick, and the rest of the Ni layer. Indeed, in the latter, we observed the twin defects extending up to the border of the interfacial layer. This layer has a rough morphology and is characterized by diffraction patterns referable to cubic and hexagonal lattices as shown in the SAED reported in **Fig. 3d**.

**Fig. 3e** shows the EDX spectrum related to the selected light blue area (~7 nm thick at the interface) of the STEM image reported as inset. On this basis, the silicided layer formed between Ni and 4H-SiC can be described as a mix of phases, *e.g* cubic $\beta$-Ni$_3$Si and hexagonal $\gamma$-Ni$_{31}$Si$_{12}$. Further details of the chemical composition in the rest of the Ni layer are shown in **Fig. 3f**. We in particular notice that the silicide layer at the interface is ~30-40 nm-thick, in agreement with what observed in Fig. 3c. Moreover, we found that the atomic concentration of Si gradually decreases towards the surface with a Ni:Si trend that is shown in **Fig. 3g** as a function of the thickness. This trend exhibits a reduced slope with respect to that of the sample laser annealed at 2.4 J/cm$^2$, and this is symptomatic of more Si diffused in the pristine nickel layer.

### 3.2.2 Identification of phases by XRD

XRD analyses were performed similarly to what done in the previous case (green line = after 3.2J/cm$^2$ irradiation; black line = reference). The treated sample maintains the texturing along the [111] direction such as in the reference sample (see **Fig. 4a** and **c**). Nonetheless, as shown in **Fig. 4b**, the main (111) peak associated to nickel is also broader (and less intense) compared to the reference case and narrower than the sample treated at 2.4 J/cm$^2$, as described by the corresponding FWHM that are: $FWHM_{(ref)}$ = 0.325°, $FWHM_{(@2.4 J/cm^2)}$ = 0.385° and $FWHM_{(@3.2 J/cm^2)}$ = 0.332°. The reduction of this parameter for the sample laser annealed at 3.2 J/cm$^2$ represents an improvement of the lattice order that could be due to the increased temperature during the laser treatment. Accordingly, an increment of the crystallite size from 24.0 nm against 21 nm estimated in the previous sample (Table 2 and Table 1). We completed the XRD analysis by collecting the rocking curve on the main (111) peak, as shown in **Fig. 4c**. Here, as observed in Fig. 2c, the sample treated shows a higher FWHM than in the reference sample. Moreover, if we focus on both treated samples, at 2.4 and at 3.2 J/cm$^2$ respectively, we found an increment of the $FWHM_{(rc)}$ (from 6.64° to 7.08°) along the main (111) peak for the sample laser annealed at 3.2 J/cm$^2$ and this represents a reduced preferential orientation of the Ni grains along the [111] direction. We further notice that the peak in Fig. 4b is asymmetric with indeed a second left-side component that is attributed to Ni+Si alloy similarly to what found in Fig. 2b. The relative area of this additional component with respect to the nickel peak is less than what found in the sample annealed at 2.4 J/cm$^2$. This finding does not imply a relative reduction of the diffused silicon atoms into the Nickel layer, since this component was also found in the grazing incidence pattern that must be indeed also considered. From the XRD analysis performed in grazing incidence configuration (see **Fig. 4d**), we found two main peaks that are slightly left shifted with respect to the reference Ni peaks (see **Fig. 4e**), and this is associated to the formation of silicide phase such as the hexagonal $\gamma$-$Ni_{31}Si_{12}$.

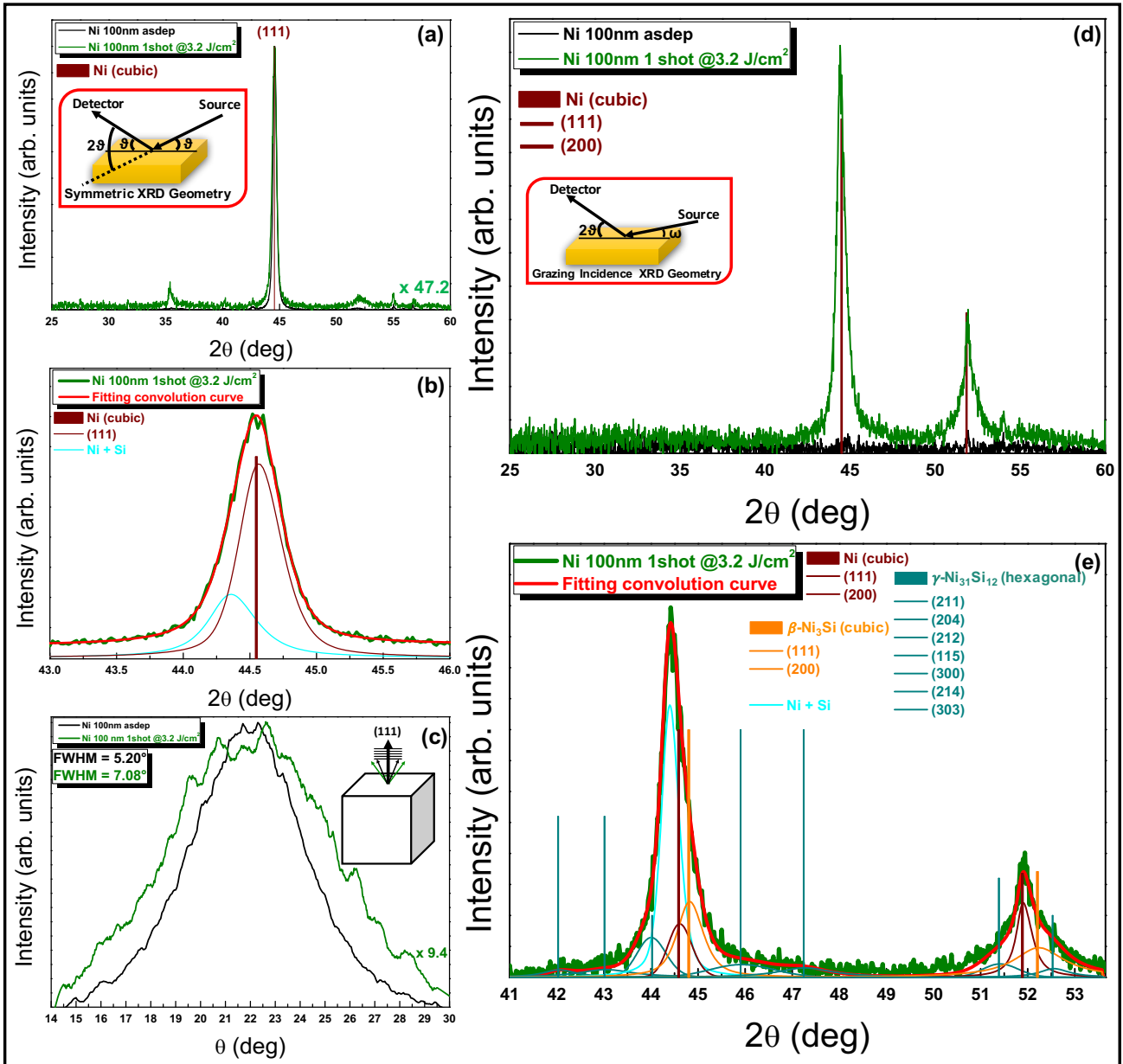

**Fig. 4**: **(a)** Normalized XRD patterns in symmetric configuration of the Ni sample laser annealed at 3.2 J/cm$^2$ for 160 ns (**green** line) and in as deposited condition (**black** line). **(b)** Magnification and deconvoluted contributions in the range around the main (111) peak and **(c)** Rocking curves of the treated vs the untreated sample. **(d)** XRD patterns in grazing incidence configuration of the layers after laser annealing at 3.2 J/cm$^2$ for 160 ns (**green** line) and in as deposited condition (**black** line). **(e)** Magnification and deconvoluted contribution of the treated sample in the range around the main (111) and (200) peaks in grazing incidence configuration. The green patterns in (a) and (c) are multiplied by a constant. Apart from pure Ni contributions, a more Si-rich phase, identified as the $\gamma$-Ni$_{31}$Si$_{12}$, is found in the pattern together with the $\beta$-Ni$_3$Si and a Ni+Si mixed alloy.

The main component under the peak at $2\vartheta \sim 44.55°$ is due to a Ni+Si alloy with $\Delta d/d = 0.44\%$ with respect to the reference Ni comment under the same peak. Other main deconvoluted components are attributed to $\gamma$-Ni$_{31}$Si$_{12}$ at $2\vartheta = 43.984°$, pure nickel at $2\vartheta = 44.588°$ and $\beta$-Ni$_3$Si at $2\vartheta = 44.802°$. This finding highlights that raising the fluence (from 2.4 to 3.2 J/cm$^2$) moves the reacted phases towards more Si-rich silicides. It has to be further noticed that the peak at $2\vartheta \sim 51.8°$ is also a convolution of different contributions, including two of the more Si-rich $\gamma$-Ni$_{31}$Si$_{12}$ phase. The peak is in all narrower than what found in the sample treated at 2.4 J/cm$^2$ and more symmetric. For the details on the peak components see table 2 (*e.g. $\beta$*-Ni$_3$Si peak: FWHM$_{(@2.4\ J/cm^2)}$ = 1.317°; FWHM$_{(@3.2\ J/cm^2)}$ = 1.200°).

**Table 2:** XRD parameters in symmetric and grazing incidence configurations related to the data in Fig. 4 (3.2 J/cm$^2$) and reporting: position (2ϑ) and interplane lattice distances (d) for each peak; Δd/d is the d variation normalised to the reference Ni peak; Full Width at Half Maximum (FWHM) of each peak; crystallite size (cry. size) for each peak component; area under each deconvoluted peak normalized to the overall area within the angular range reported in column 3 (realtive area %). An instrumental discrepancy between symmetric and grazing peak position of 0.04° is accounted in the reported values. A nanostructured hexagonal $\gamma$-Ni$_{31}$Si$_{12}$ phase is additionally formed after laser annealing.

| XRD Geometry | Phase | Peak range (in 2ϑ) | Plane | Position 2ϑ (°) | Position d (Å) | Δd/d (%) | FWHM (°) | Cry. Size (nm) | Area Relative | Area Total | Relative area (%) | Resume | |
|---|---|---|---|---|---|---|---|---|---|---|---|---|---|
| **Symmetric** | Ni + Si | 43° - 46° | - | 44.340 | 2.042 | 0.45 | 0.372 | 22 | 0.4 | | 29.2 | | |
| | Ni (cubic) | | (111) | 44.548 | 2.033 | - | 0.332 | 24 | 1.0 | | 70.8 | | |
| | | | | | | | | | | | | **Resume** | |
| **Grazing Incidence** | $\gamma$-Ni$_{31}$Si$_{12}$ (hexagonal) | 41° - 47.5° | (211) | 42.029 | 2.149 | - | 0.800 | 10 | 2.0 | 78 | 2.6 | Ni + Si | 42.6 |
| | | | (204) | 43.013 | 2.102 | - | 1.300 | 6.1 | 3.5 | | 4.5 | Ni (cubic) | 9.7 |
| | | | (212) | 43.984 | 2.058 | - | 0.670 | 12 | 6.5 | | 8.4 | $\beta$-Ni$_3$Si (cubic) | 19.4 |
| | Ni + Si | | - | 44.380 | 2.040 | 0.44 | 0.400 | 20 | 33 | | 42.6 | $\gamma$-Ni$_{31}$Si$_{12}$ (hexagonal) | 28.4 |
| | Ni (cubic) | | (111) | 44.588 | 2.031 | - | 0.550 | 15 | 7.5 | | 9.7 | | |
| | $\beta$-Ni$_3$Si (cubic) | | (111) | 44.802 | 2.022 | - | 0.600 | 13 | 15 | | 19.4 | | |
| | $\gamma$-Ni$_{31}$Si$_{12}$ (hexagonal) | | (115) | 45.903 | 1.976 | - | 1.500 | 5.3 | 5.0 | | 6.5 | | |
| | | | (300) | 47.242 | 1.923 | - | 1.300 | 6.1 | 5.0 | | 6.5 | | |
| | $\gamma$-Ni$_{31}$Si$_{12}$ (hexagonal) | 50.5° - 53.5° | (214) | 51.387 | 1.777 | - | 0.840 | 9.5 | 3.6 | 36 | 9.9 | Ni (cubic) | 34.5 |
| | Ni (cubic) | | (200) | 51.887 | 1.762 | - | 0.321 | 25 | 12 | | 34.5 | $\beta$-Ni$_3$Si (cubic) | 47.3 |
| | $\beta$-Ni$_3$Si (cubic) | | (200) | 52.205 | 1.752 | - | 1.200 | 6.6 | 17 | | 47.3 | $\gamma$-Ni$_{31}$Si$_{12}$ (hexagonal) | 18.3 |
| | $\gamma$-Ni$_{31}$Si$_{12}$ (hexagonal) | | (303) | 52.528 | 1.742 | - | 0.700 | 11 | 3.0 | | 8.3 | | |

### 3.3 100 nm-Ni/4H-SiC annealed @3.8 J/cm$^2$

#### 3.3.1 Structural and compositional analysis by TEM and EDX

The systematic characterizations performed for the previous samples are carried out for the sample 1 shot laser annealed at 3.8 J/cm$^2$ for 160 ns. In this latter case, after LA we found that the thickness of the Ni layer further increased by 67 nm due to augmented interfacial reaction, as will be discussed hereafter.

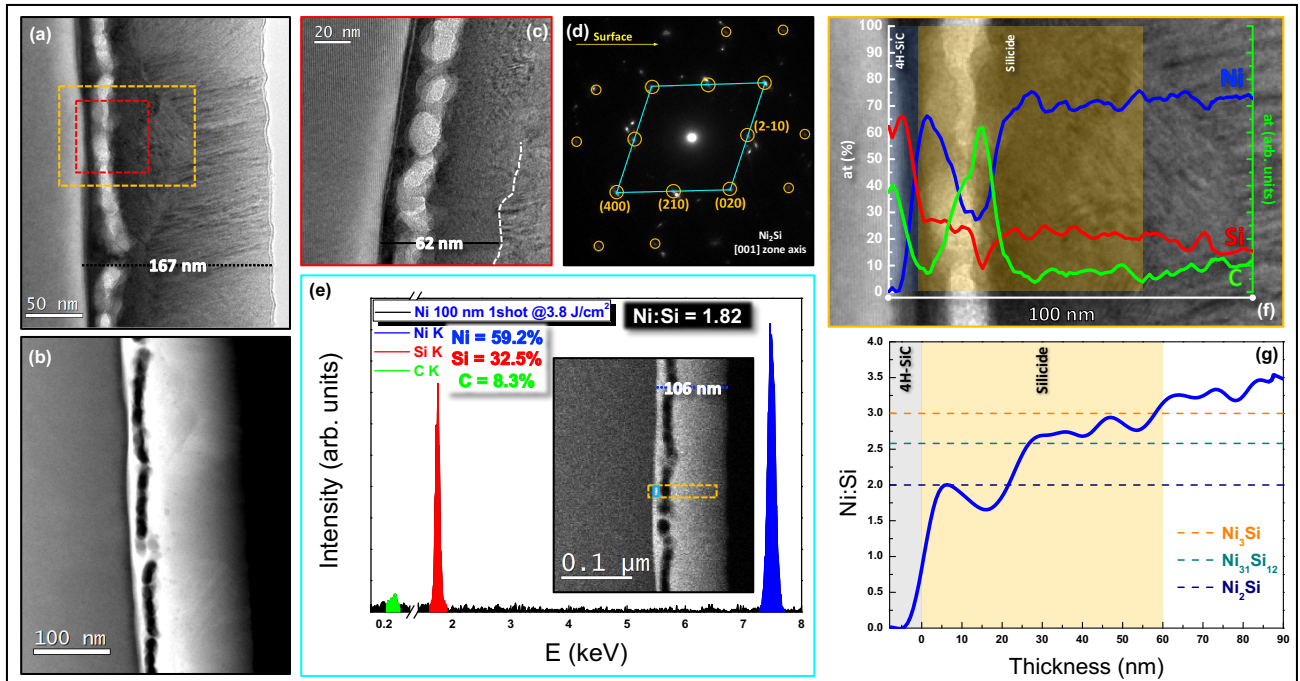

**Fig. 5:** cross-sectional TEM images **(a)** in bright field and **(b)** in dark field of the Ni sample (167 nm) laser annealed 1 shot at 3.8 J/cm$^2$ for 160 ns. **(c)** Magnification of the red panel in the Fig. (a) relative a thin part of the Ni layer in order to show the difference of the Ni layer at the interface with the substrate. **(d)** SAED image taken at the interfacial layer that testifies local silicide reaction in the $\delta$-Ni$_2$Si phase. **(e)** EDX spectrum of the light blue panel in the STEM image in inset. **(f)** EDX atomic distribution profile in the region inside the yellow panel in the Fig. (a) and in inset of Fig. (e) through the layer. **(g)** Ni:Si ratio profile from the interface with 4H-SiC through the entire layer. An interfacial layer, 55-60 nm-thick, is formed with silicide composition expected between $\beta$-Ni$_3$Si and $\delta$-Ni$_2$Si as indicated by the dashed lines in the Fig. Above this interface, Si atoms diffuse into the Ni layer as shown by the blue curve. C atoms accumulation above the silicide/4H-SiC interface produces a deepening of Ni:Si curve at around 15 nm. The corresponding C peak is shown by the green line in Fig. (f).

From the morphology point of view and in analogy to what observed in all the other samples, most of the layer is made of big grains with defects such as inclusions and twins as shown in **Fig. 5a**. From the TEM image in dark field (see **Fig. 5b**) we found that the C-clusters further increased in number and size, indeed, they formed a continuous layer, ~20 nm-thick, that is 10-15 nm away from the 4H-SiC substrate even if some of them are localized up to ~30 nm from the interface. The magnification shown in **Fig. 5c** is relative to the red panel selected in Fig. 5a. Here, we observe with more detail the difference in morphology between the interfacial layer with 4H-SiC, ~50-60 nm thick, and the rest of the Ni layer. From the structural point of view and differently from the silicided layer of the previous samples, typical diffraction patterns of the orthorhombic $\delta$-Ni$_2$Si phase are found (**Fig. 5d**) together with the cubic and hexagonal phases already formed at lower fluences. Mixed composition in the silicided layer is further confirmed by local EDX analyses and related profiles (**Fig. 5e and 5f).** We notice that the silicide layer thickness, ~50-60 nm, agrees with what observed in the figure 5c. In the rest of the layer, Si atoms have diffused into the nickel matrix gradually decreasing towards the surface. **Fig. 5g** shows the trend of the Ni:Si ratio as a function of the thickness with a reduced slope with respect to the previous cases.

### 3.3.2. Identification of phases by XRD

XRD analyses in the treated and reference samples are shown in **Fig. 6** (blue and black line, respectively). Even in this case, the residual part of nickel maintains the texturing along the [111] direction as in the reference sample (see **Fig. 6a**). As shown in **Fig. 6b**, the main (111) peak is similar to the reference case but less intense and narrower than the corresponding peaks in the samples treated at lower fluences (see Fig. 2b and 4b). The FWHM in the reference and in this treated sample are indeed comparable, being $FWHM_{(ref)}$ = 0.325° and $FWHM_{(@3.8\ J/cm^2)}$ = 0.328°. Instead, the reduction of the FWHM compared to the rest of annealed samples is symptomatic of an improvement of the crystallographic quality in the residual topmost Ni layer (see Fig. 5a). Moreover, this testifies a progressive increment of the crystallite size, being 25 nm against 24 nm and 21 nm estimated in the samples annealed at 3.2 and 2.4 J/cm$^2$, respectively. The process of symmetrization of the (111) nickel peak, that is systematic while increasing the fluence (see Figs. 2b, 4b, 6b), is ascribed to a progressive consumption of the nickel layer from interfacial reactions that removes a portion of the nickel layer that has been primarily affected by the thermal field and likely by Si diffusion. This explains why, only at the highest fluence, the residual nickel peak (we recall that it is less intense than the one in the reference) approaches the one before annealing although the Ni+Si peak is not any more visible under the (111) Ni peak, a contribution from the alloy can be identified in grazing incidence pattern. The rocking curve on the (111) peak is shown in **Fig. 6c**. Differently from the other cases, the peak is left shifted by ~2° and this demonstrates that the preferential growth axis for those grains is twisted. In addition, we found a reduction of the FWHM$_{(rc)}$ (0.566°) for this sample against the other treated samples and this means that at high fluence the structural order is increased.

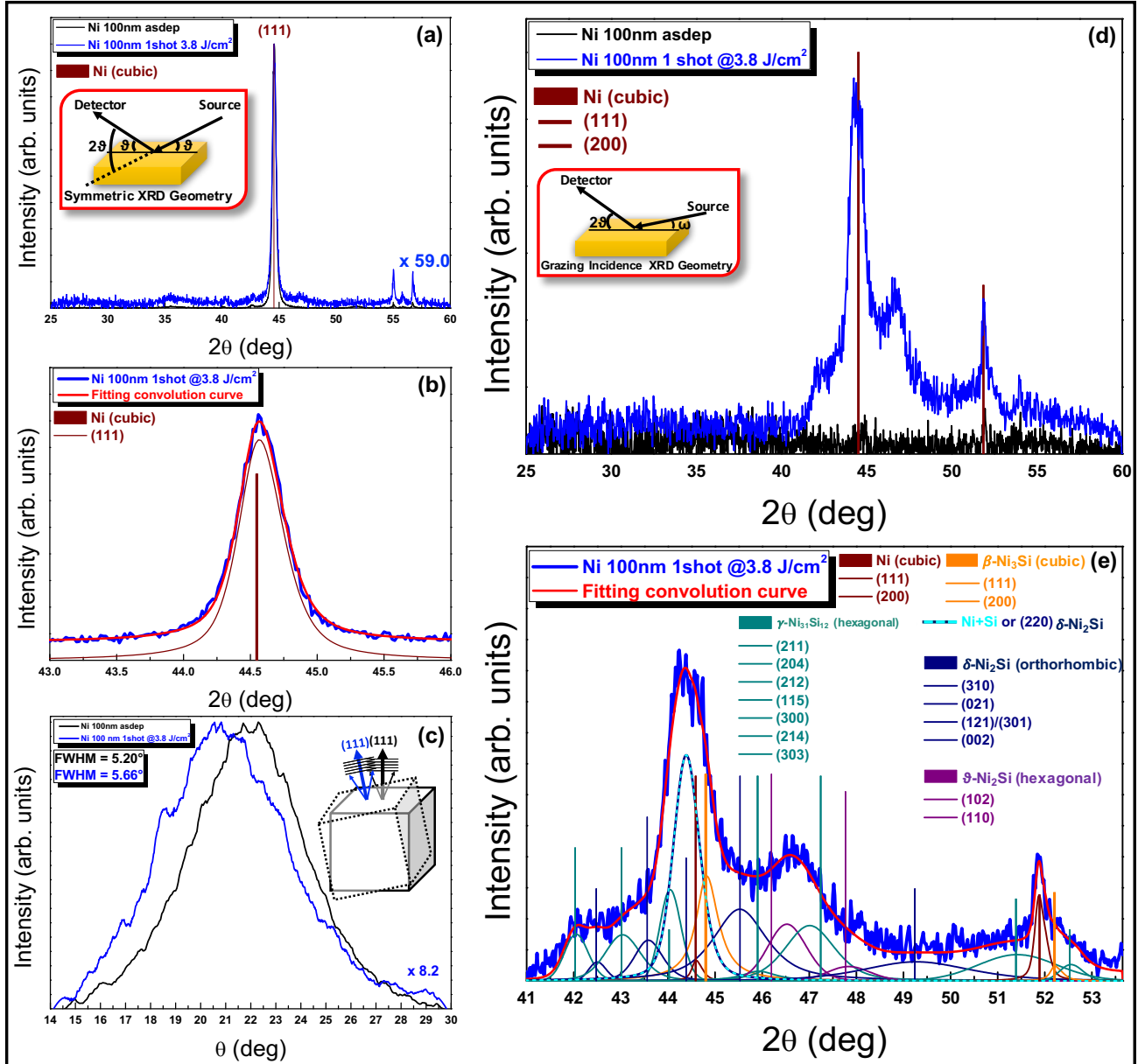

**Fig. 6**: **(a)** normalized XRD patterns in symmetric configuration of the Ni sample laser annealed at 3.8 J/cm$^2$ for 160 ns (**blue** line) and in as deposited condition (**black** line). **(b)** Magnification and deconvoluted contributions in the range around the main (111) peak and **(c)** Rocking curves of the treated vs the untreated sample **(d)** XRD patterns in grazing incidence configuration of the layers after laser annealing at 3.8 J/cm$^2$ for 160 ns (**blue** line) and in as deposited condition (**black** line). **(e)** Magnification and deconvoluted contribution of the treated sample in the main range in grazing incidence configuration. We note that the broad peak in the range 46-47.5° can be equivalently interpreted by two main peaks of the Ni$_{74}$Si$_{26}$ or Ni$_{25}$Si$_9$ metastable phases [34]. The blue patterns in (a) and (c) are multiplied by a constant. The vertical bars in (e) are reference markers of the main peak in the respective phases. Apart from pure Ni contributions, more Si-rich phases identified as the $\vartheta$–Ni$_2$Si and $\delta$-Ni$_2$Si are found in the pattern together with the $\gamma$–Ni$_{31}$Si$_{12}$, $\beta$-Ni$_3$Si and a Ni+Si mixed alloy.

In grazing incidence configuration (see **Fig. 6d**) we found an intriguing scenario with several contributions in the pattern that denote the formation of multiple silicide phases. In particular, the 2ϑ range between 41° and 48° can be deconvoluted into component peaks as shown in **Fig. 6e**.

**Table 3:** XRD parameters in symmetric and grazing incidence configurations related to the data in Fig. 6 (3.8 J/cm$^2$) and reporting: position (2ϑ) and interplane lattice distances (d) for each peak; Δd/d is the d variation normalised to the reference Ni peak; Full Width at Half Maximum (FWHM) of each peak; crystallite size (cry. size) for each peak component; area under each deconvoluted peak normalized to the overall area within the angular range reported in column 3 (realtive area %). An instrumental discrepancy between symmetric and grazing peak position of 0.04° is accounted in the reported values.

| XRD Geometry | Phase | Peak range (in 2ϑ) | Plane | Position 2ϑ (°) | Position d (Å) | Δd/d (%) | FWHM (°) | Cry. Size (nm) | Area Relative | Area Total | Relative area (%) | | |
|---|---|---|---|---|---|---|---|---|---|---|---|---|---|
| Symmetric | Ni (cubic) | 43° - 46° | (111) | 44.548 | 2.033 | - | 0.328 | 25 | 60 | | 100.0 | | |
| | | | | | | | | | | | **Resume** | | |
| Grazing Incidence | γ-Ni$_{31}$Si$_{12}$ (hexagonal) | 41° - 48° | (211) | 42.029 | 2.149 | - | 0.551 | 15 | 2.9 | 84 | 3.5 | Ni (cubic) | 1.0 |
| | δ-Ni$_2$Si (orthorhombic) | | (310) | 42.478 | 2.127 | - | 0.400 | 20 | 0.9 | | 1.1 | β-Ni$_3$Si (cubic) | 11.9 |
| | γ-Ni$_{31}$Si$_{12}$ (hexagonal) | | (204) | 43.013 | 2.102 | - | 0.832 | 10 | 5.1 | | 6.0 | γ-Ni$_{31}$Si$_{12}$ (hexagonal) | 29.7 |
| | δ-Ni$_2$Si (orthorhombic) | | (021) | 43.561 | 2.077 | - | 0.651 | 12 | 3.6 | | 4.3 | | |
| | γ-Ni$_{31}$Si$_{12}$ (hexagonal) | | (212) | 43.013 | 2.102 | - | 0.483 | 17 | 5.4 | | 6.5 | | |
| | Ni+Si or δ-Ni$_2$Si | | (220) | 44.365 | 2.041 | 0.04 | 0.609 | 13 | 20 | | 23.3 | Ni+Si or δ-Ni$_2$Si | 23.3 |
| | Ni (cubic) | | (111) | 44.588 | 2.031 | - | 0.200 | 42 | 0.8 | | 1.0 | | |
| | β-Ni$_3$Si (cubic) | | (111) | 44.802 | 2.022 | - | 0.502 | 16 | 10 | | 11.9 | δ-Ni$_2$Si (orthorhombic) | 23.3 |
| | δ-Ni$_2$Si (orthorhombic) | | (121)/(301) | 45.499 | 1.993 | 0.06 | 1.250 | 6.4 | 15 | | 17.9 | | |
| | γ-Ni$_{31}$Si$_{12}$ (hexagonal) | | (115) | 45.903 | 1.976 | - | 0.800 | 10 | 1.5 | | 1.8 | | |
| | ϑ-Ni$_2$Si (hexagonal) | | (102) | 46.496 | 1.952 | -0.53 | 0.900 | 8.9 | 7.0 | | 8.4 | ϑ-Ni$_2$Si (hexagonal) | 10.8 |
| | γ-Ni$_{31}$Si$_{12}$ (hexagonal) | | (300) | 47.002 | 1.933 | 0.48 | 1.150 | 6.9 | 10 | | 11.9 | | |
| | ϑ-Ni$_2$Si (hexagonal) | | (110) | 47.808 | 1.902 | - | 1.000 | 8.0 | 2.0 | | 2.4 | | |
| | δ-Ni$_2$Si (orthorhombic) | 48° - 53.5° | (311) | 49.241 | 1.850 | - | 2.500 | 3.2 | 8.4 | 23 | 37.1 | Ni (cubic) | 18.8 |
| | γ-Ni$_{31}$Si$_{12}$ (hexagonal) | | (214) | 51.387 | 1.777 | - | 0.840 | 4.3 | 7.6 | | 33.7 | β-Ni$_3$Si (cubic) | 4.0 |
| | Ni (cubic) | | (200) | 51.887 | 1.762 | - | 0.321 | 42 | 4.2 | | 18.8 | γ-Ni$_{31}$Si$_{12}$ (hexagonal) | 40.1 |
| | β-Ni$_3$Si (cubic) | | (200) | 52.205 | 1.752 | - | 1.200 | 38 | 0.9 | | 4.0 | | |
| | γ-Ni$_{31}$Si$_{12}$ (hexagonal) | | (303) | 52.528 | 1.742 | - | 0.700 | 15 | 1.4 | | 6.4 | δ-Ni$_2$Si (orthorhombic) | 37.1 |

Pivotal information can be extracted as follows: 1) the new evidence on the formation of the orthorhombic δ-Ni$_2$Si, in agreement with Fig. 5d, and of the hexagonal ϑ-Ni$_2$Si phases with respect to the other analyzed cases; 2) a peak of the Ni+Si alloy can be found at 2ϑ = 44.365°, even though this could be equally attributed to slightly strained (by 0.04%) (220) planes of the δ-Ni$_2$Si phase (both attributions are labelled in Fig. 6e); 3) hexagonal phases such as $\gamma$-$Ni_{31}Si_{12}$ are found similarly to what found in the 3.2 J/cm$^2$ treated samples; 4) the broad peak in the range 46-47.5° has been represented by a double peak deconvolution using two strained (102) $\vartheta$-$Ni_2Si$ and (300) $\gamma$-$Ni_{31}Si_{12}$ peaks; it could be equivalently interpreted by unstrained (1,1,12) and (0,3,0) peaks of the hexagonal $Ni_{74}Si_{26}$ (or $Ni_{25}Si_9$) phase, that is given in the literature as a metastable phase observed after fast cooling from a Ni:Si $\sim$ 2.9 melted alloy [25,34]; 5) $\beta$-$Ni_3Si$ contributions are also found. Another main feature in the diffraction pattern is found in the range $2\vartheta \sim$ 48-53.5° in which main contributors are $\gamma$-$Ni_{31}Si_{12}$, $\delta$-$Ni_2Si$, pure Ni and $\beta$-$Ni_3Si$ phases. The overall scenario corroborates the trend towards the formation of more Si-rich silicide phases, such as the $\delta$-$Ni_2Si$ and the $\vartheta$-$Ni_2Si$, when increasing the laser fluence. Most importantly, the $\delta$-$Ni_2Si$ phase can be localized at the interface with the 4H-SiC substrate on the basis of TEM results such as the one shown in Fig. 5c/d and also in agreement with the Ni/Si chemical profile in Fig. 5g.

## 4. Phase field simulations

Pulsed laser processes activate a complex ultrafast material evolution characterized by different stages and kinetic pathways when process parameters vary. As a consequence, the characterizations' evidences, discussed in the section 3, are relative to the final result of this evolution, whereas the time dependent details are beyond the experimental analysis only. In order to gain a complete scenario, the LA processes have been simulated using, as input, the same initial system (*i.e.* 100nm thick Ni layer on 110 µm thick 4H-SiC substrates) and the same process parameters (*i.e.* pulses shapes and fluences).

The evolution framework derived from the simulations can be qualitatively separated in three main regimes: sub-melting, partial-melting, full-melting characterized by different features [32]. Hereby we summarize these features:

a) In the sub-melting regime, *i.e.* the low fluence range, the systems remain solid during the whole heating/cooling cycle and the main phenomena (both boosted by the temperature increase) are the intermixing, which starts at the Ni/4H-SiC interface, and the subsequent early and limited silicide formations if the Si density is above solid solubility threshold [32].

b) In the partial melting regime, a limited region of the Ni-rich layer melts, whilst the rest of the layer remains solid, after an initial intermixing and silicide formation due to the temperature increases. The solid portion follows during the processes an evolution similar to the sub-melting case whilst the molten region deviates from this evolution in the melting transient stage, as explained in the following. In all the investigated cases, the molten phase is almost ideal liquid Ni-Si-C alloy with the eventual co-presence of solid carbon (C-cluster). Consequently, during the melting stage silicide compounds, formed during the heating (see case a), locally release Si and Ni in the monomer forms. The intermixing of the molten region is very strong due to the high diffusivity of the monomer species in the liquid. After the solidification, while the temperature is still high, the formation of silicide restarts, but from an intermixed state completely different from the one achieved in the solid portion.

c) In the full melting regime (higher fluences) the whole Ni-rich layer melts (from the surface to the 4H-SiC substrate). It is obvious that in this case the kinetic discussed in the previous item for the molten region extends with similar characteristics to the whole Ni-rich film.

The three processes in study can be classified as it follows: 2.4J/cm$^2$ case b), 3.2 J/cm$^2$ and 3.8 J/cm$^2$ case c). The melting stage starts (ends) after 133ns (228ns), 103ns (292ns), 86ns (340ns) for the 2.4J/cm$^2$, 3.2 and 3.8 J/cm$^2$ fluences respectively. The maximum molten region is 19.1 nm, 132.7 and 140.6nm again for the 2.4J/cm$^2$, 3.2 4J/cm$^2$ and 3.8 J/cm$^2$ cases. We note that the melt depth can extend beyond the original Ni-SiC interface since the interphase mixing decreases locally the melting point with respect to the 3100K value of the pure 4H-SiC material [35,36].

An interesting aspect of the simulated melting phenomena is that it nucleates at the Ni-SiC interface for the 2.4J/cm$^2$ fluence and at the surface with air for the other two cases. This peculiar feature is caused by the trade-off between the intermixing evolution which tends to lower the composition dependent $T_m[X_A]$, melting point of the ternary system and the induced temperature gradient, typical of the laser process, where the maximum temperature is achieved at the surface. For example the binary phase diagram of the Ni-Si system, in the pure Ni the melting temperature is 1728 K while it reaches the value of 1424K for a Si:Ni stoichiometric ratio of about 0.2, and a similar trend is observed in the Ni-Si-C system [36].

In order to understand the reasons for a different location of the initial melting we should consider an ideal snapshot of the temperature and composition fields just before the melting. We observe two gradients: a) one of the temperature field $T(x)$ where the T maximum occurs at the surface; b) another of the local melting point $T_m[X_A(x)]$, again with the maximum at the surface, since $T_m(X_A)$ follows the Ni element profile which decreases monotonically from the surface to the SiC interface. The melting nucleates when $T(x^*) > T_m[X_A(x^*)]$ and $x^*$ could be either the surface $x^* = 0$ or close to the Ni-SiC interface according the actual $T(x^*)$ and $T_m[X_A(x^*)]$ profiles which can strongly depend on the fluence.

The temperature drops between the surface and original Ni-SiC interface at the melting onset is ΔT=83K (from 1542K to 1459K), 115K (from 1729K to 1614K), 132K (from 1730K to 1598K) for the LA at the 2.4J/cm$^2$, 3.2 and 3.8 J/cm$^2$ fluences respectively. Consistently to the previous consideration in the 2.4J/cm$^2$ case the strong intermixing close to the SiC (Si:Ni=0.22) interface favors the nucleation of the melting in this position in spite to the lower temperature, whilst in the other case the intermixing in not sufficient to balance the temperature difference and the nucleation occurs at the (hottest) surface zone.

In fig. 7(a-c) the local density profile of silicide compounds obtained by means of the laser annealing simulations of the processes at the three fluences. We show only the $Ni_3Si$, $Ni_{31}Si_{12}$ and $Ni_2Si$ since the density of other compounds is less significant (less than 0.01 in the used scale). We note that the local fraction of the Ni and Si elements in the compounds can be obtained from the plotted quantity simply multiplying this quantity by the respective stoichiometry coefficients.

The 2.4J/cm$^2$ fig. 7 a) is classified as partial melting case and the silicide region practically coincides with the molten region (about 20nm thick). The simulations evidence that, in this and the other cases, the ultra-fast mixing occurring in the liquid state is a crucial boosting mechanism for the silicide formation; indeed, *e.g.* the $Ni_3Si$ density maximum is below the 0.03 level before the melting onset at 2.4J/cm$^2$ fluence.

The $Ni_3Si$ compound is assumed to start Ni-silicide reaction chain and it is the first compound which forms at the Ni-SiC interface, *e.g.* in Ni diffuse tail in the SiC region. As a consequence, a sharp peak of the $Ni_3Si$ density is present in this and in the other cases at the interface with the SiC material. This peak dominates the silicide region in the 2.4J/cm$^2$ case and the $Ni_3Si$ represents the majority silicide compound for this case, while the intensity of the other two components is significantly less important. Moreover, also the depth of the SiC interface as a function of the process could be easily visualized by these peak: it is located at ≈ 122nm, 132nm, 140nm for the 2.4J/cm$^2$, 3.2 and 3.8 J/cm$^2$ fluences respectively. In the latter two cases it is, obviously, consistent with the maximum molten depth. We note that in order to correctly estimate the Ni+Ni-silicide layer thickness (i.e. the Ni-rich layer in the form of a solid solution of the three elements + the silicide layer at the interface with the SiC substrate) from these values we need add a correction due to the volume modification (not considered in the phase field model) related to the different atomic density of the compounds with respect to the ones of the pristine material. Anyhow, considering this correction, the thicknesses of the Ni+Ni-silicide films estimated by the simulations are in a nice agreement with the experimental measurements.

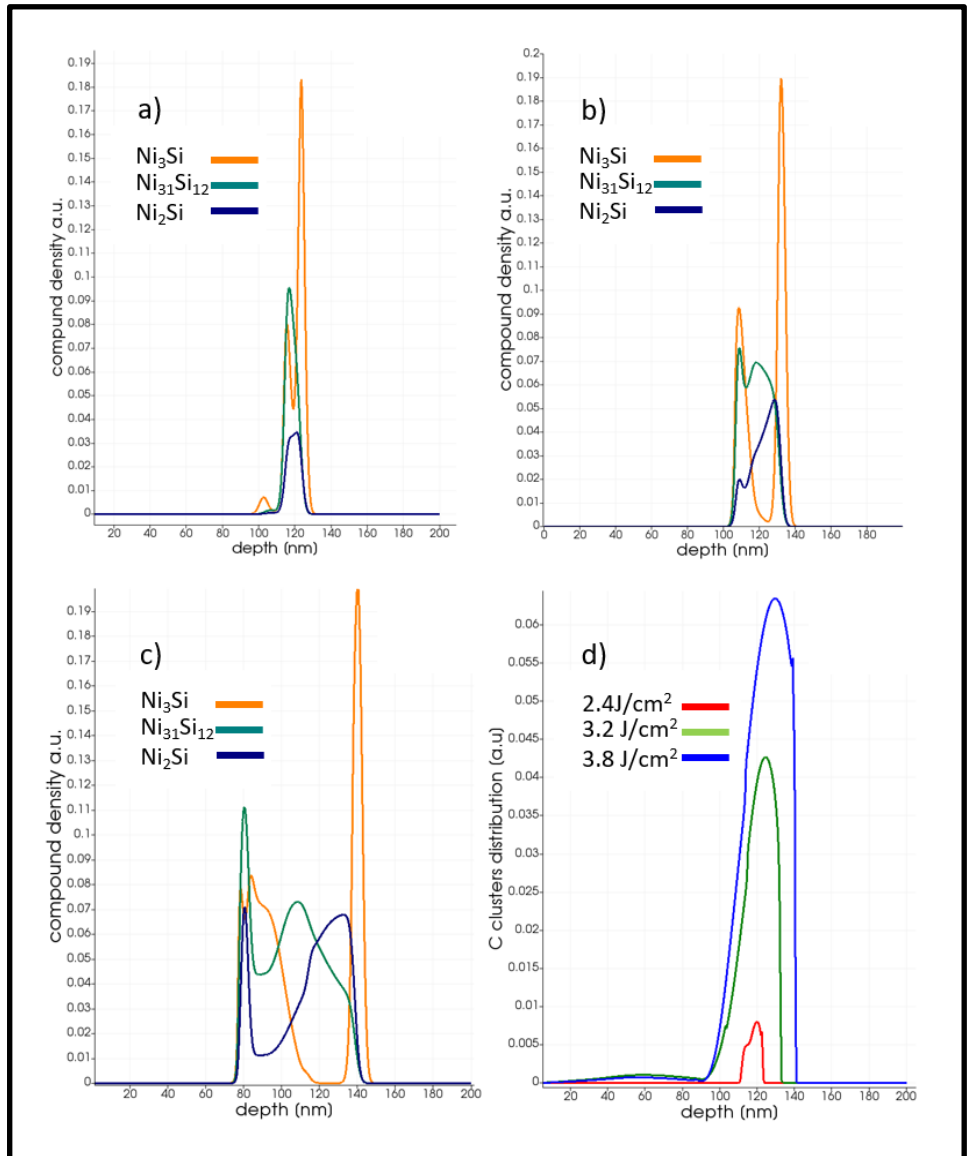

**Fig. 7**: Simulated Local density of the $Ni_3Si$ (red lines), $Ni_{31}Si_{12}$ (green lines) and $Ni_2Si$ (blue lines) for Laser Annealing processes with fluences 2.4J/cm$^2$ **(a)**, 3.2 J/cm$^2$ **(b)** and 3.8 J/cm$^2$ **(c)**. **(d)** Local carbon cluster space distribution as a function of the laser annealing fluence. The cases shown are 2.4J/cm$^2$ (light blue line), 3.2 J/cm$^2$ (magenta line) and 3.8 J/cm$^2$ (black line).

Analogously, a good agreement between simulations and experiments is obtained for the estimates of the Ni-silicide region thickness: the simulations predict ≈20nm, 40nm, 60nm thick Ni-silicide films for the 2.4J/cm$^2$, 3.2 J/cm$^2$ and 3.8 J/cm$^2$ fluences respectively. We can also derive that, in spite the full melting classification of the 3.2 J/cm$^2$ and 3.8 J/cm$^2$ cases, the silicide region is still confined and adjacent to the SiC interface as experimentally evidenced: the intermixing entity and residual thermal budget in the post-melting stage (both increasing with the fluence) are again the two driving forces for its formations. Regarding the balance between the silicide components, simulations recover the increasing relevance of Si-richer compound as the fluence increases: for the 3.2 J/cm$^2$ the $Ni_3Si$ and $Ni_{31}Si_{12}$ have similar weights in the silicide layer while the $Ni_2Si$ material seems less relevant (Fig. 7. b). Finally, for the 3.8 J/cm$^2$ case the three compounds characterize *au pair* the silicide region (fig. 7c).

The latter phenomenon characterizing the kinetic evolution during the irradiation is the clustering of C atoms which are locally in a supersaturation level with respect the 1:1 SiC stoichiometry. In fig. 7d the relative final density of C-cluster is reported as function of the fluence for the three studied cases. As consequence of the rapid diffusion of the C monomers the average clustering rate is very high in the melting stage. We note that a fraction of solid C could be stable in the Ni-Si liquid due to its high melting point, *i.e.* C monomers segregate in the liquid. The height and extension of the C-cluster profiles strongly depends on the fluence in a noteworthy qualitative agreement with the TEM analysis.

**Conclusion**

In conclusion, we analyzed three Ni samples that were laser-annealed at different fluences by single shot at 2.4, 3.2 and 3.8 J/cm$^2$, in order to investigate silicide phases formation after the LA treatments. From the cross-correlated characterizations, we conclude a common behavior with the reaction of a silicide layer at the interface and C-clusters that are located 5-15 nm away from the interface with 4H-SiC. Both proportionally change as a function of the fluence. In particular, at the lowest fluence, the C-clusters are sporadic and small whilst they increase in size and number by increasing the fluence. In addition, another common finding is the topmost layer that is made of big grains with defects such as inclusions and twins extending up to the interfacial silicide layer. The thickness of this layer reduces with increasing fluence.

**Fig. 8** shows a qualitative schematic highlighting the silicide phases formed starting from the interface with 4H-SiC, based on the compositional/structural findings, and silicon atomic diffusion into the pristine nickel layer. On the basis of XRD results we argue that progressively more silicon-rich phases are formed by increasing fluence. As a matter of fact, $Ni_2Si$ is specifically found at high fluences (3.8J/cm$^2$). Moreover, under-stoichiometric Si mixed to Ni create a topmost layer with lattice distortion of the host cubic Ni lattice. This reflects in specific features (FWHM, peak position, broadening) of XRD patterns.

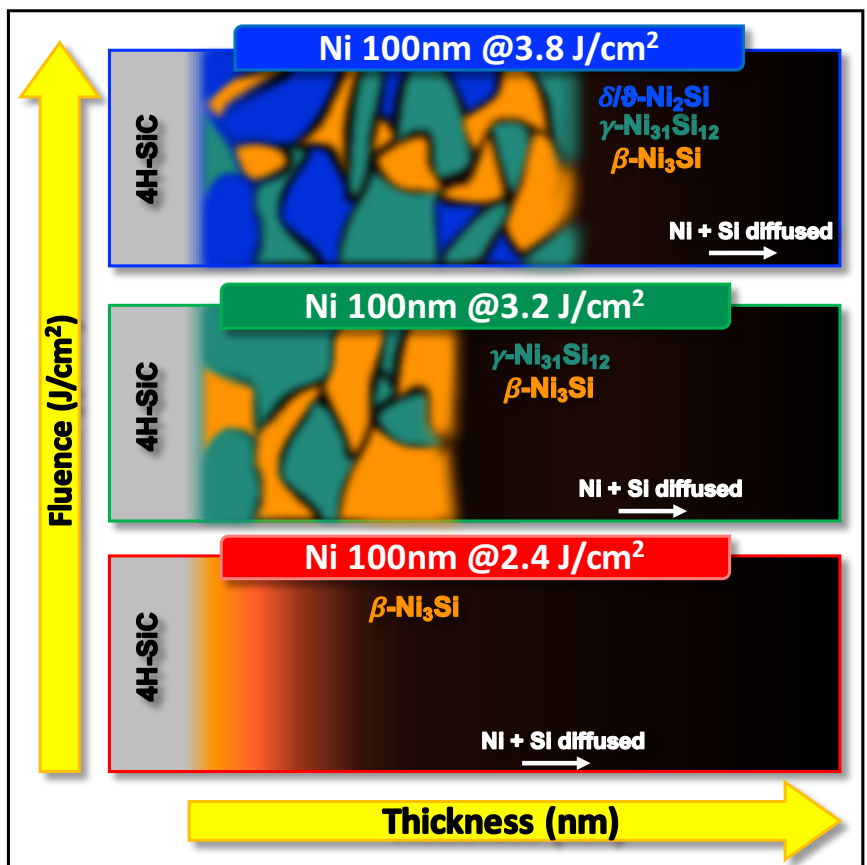

**Fig. 8**: Qualitative schematic of the three 1 shot laser annealed Ni samples at different fluence, **2.4**, **3.2** and **3.8** J/cm$^2$ respectively. The different silicide phases are gradually shown in colors: **orange** cubic $\beta$-$Ni_3Si$ phase, **dark-cyan** hexagonal $\gamma$-$Ni_{31}Si_{12}$ phase, **blue navy** orthorhombic $\delta$-$Ni_2Si$ phase and **brown** degrading to **black** a mixed Ni+Si layer with Si progressively diffusing towards the surface of the sample.

Pulsed LA activate a complex ultrafast material evolution characterized by different stages and kinetic pathways when process parameters vary that were studied by Face Field Simulations. The qualitative evolution framework derived from the simulations can be separated in three main regimes: sub-melting, partial-melting and full-melting. The simulated scenario, according to the experimental findings, provides an important tool to describe the phenomena occurring during LA with further predictive valence at varied boundary conditions.

## ACKNOWLEDGEMENTS

The authors thank the project **MADEin4** (Metrology Advances for Digitized Electronic Components and Systems Industry 4.0) that has received funding from the ECSEL JU under grant agreement No 826589. The JU receives support from the European Union's Horizon 2020 research and innovation programme and France, Germany, Austria, Italy, Sweden, Netherlands, Belgium, Hungary, Romania and Israel.